\documentclass[prl,showpacs,twocolumn]{revtex4}

\usepackage{graphicx}
\usepackage{bm}
\usepackage{pifont}
\begin{document}

\title{Impurity-doping induced ferroelectricity in frustrated antiferromagnet CuFeO$_2$}

\author{S. Seki$^1$, Y. Yamasaki$^1$, Y. Shiomi$^1$, S. Iguchi$^1$, Y. Onose$^1$, and Y. Tokura$^{1,2,3}$} 
\affiliation{$^1$ Department of Applied Physics, University of Tokyo, Tokyo 113-8656, Japan \\ $^2$  Spin Superstructure Project, ERATO, Japan Science and Technology Agency (JST), Tsukuba 305-8562, Japan \\ $^3$ Correlated Electron Research Center (CERC), National Institute of Advanced Industrial Science and Technology (AIST), Tsukuba 305-8562, Japan }

\date{November 9, 2006}

\begin{abstract}

Dielectric responses have been investigated on the triangular-lattice antiferromagnet CuFeO$_2$ and its site-diluted analogs CuFe$_{1-x}$Al$_x$O$_2$ ($x$=0.01 and 0.02) with and without application of magnetic field. We have found a ferroelectric behavior at zero magnetic field for $x$=0.02. At any doping level, the onset field of the ferroelectricity always coincides with that of the noncollinear magnetic structure while the transition field dramatically decreases to zero field with Al doping. The results imply the further possibility of producing the ferroelectricity by modifying the frustrated spin structure in terms of site-doping and external magnetic field. 

\end{abstract}
\maketitle

The ferroelectricity in magnetic materials has been attracting much attention because of possibility of showing gigantic magneto-electric effects\cite{Review1, Review2, Revival}, as stimulated by the discovery of the electric polarization flop induced by a magnetic field in TbMnO$_3$\cite{TbMn}. Similar ferroelectric behaviors have recently been discovered in several spin-frustrated systems; Ni$_3$V$_2$O$_8$\cite{Ni3V2O8}, Ba$_{0.5}$Sr$_{1.5}$Zn$_2$Fe$_{12}$O$_{22}$\cite{plumbite}, CoCr$_{2}$O$_4$\cite{CoCr2O4}, and MnWO$_4$\cite{MnWO4}. In all these materials, the ferroelectric behavior is observed in the noncollinear magnetic phase. Katsura {\it et al.} proposed the spin current theory for the origin of ferroelectricity in the noncollinear magnetic phase\cite{Katsura}. In the theory, the electric dipole is induced by the spin current, which is expected to flow between noncollinear magnetic moments in analogy to the magnetic dipole induced by electric current. To produce such a ferroelectricity of magnetic origin, modification or partial lifting of spin state degeneracy may be prerequisite with use of the frustrated spin systems.

Delafossite CuFeO$_2$ (See Fig. 1(d) for the structure) is one of candidates for such magnetic ferroelectrics. The crystal structure is characterized by the space group $R\bar{3}m$, with two-dimensional triangular lattice layers stacked rhombohedrally along the $c$-axis (Fig. 1(d)). The magnetic structure of this material has been studied extensively by neutron diffraction measurements\cite{PD_structure, x000_FI, x002, zero_field_diagram, OPD_structure, ODSx002}. At zero field, a 4-sublattice collinear structure (CM-4) is observed at the lowest temperature. When the field is increased up to 7 T, a noncollinear spin structure (NC) emerges\cite{meeting}. Above 12 T, a 5-sublattice collinear structure (CM-5) is realized\cite{x000_FI}. Recently, Kimura {\it et al.} reported that the spontaneous polarization is observed in the noncollinear magnetic phase in between the 4-sublattice and 5-sublattice phases\cite{Kimura}. Similarly to the magnetic field, Al doping is known to easily modify the magnetic structure\cite{zero_field_diagram}. The 2 \% substitution of Fe with Al induces the transition from CM-4 to NC at zero field and low temperature\cite{x002}. To clarify the relation between the magnetic structure and dielectric properties, the search for electric polarization is desirable also for the Al-doped crystals. In this work, we report the finding of ferroelectricity in the noncollinear magnetic phase of the Al doped crystal even in the absence of external magnetic field as well as the systematic evolution of the magneto-electric phase with Al-doping. This ensures the close relation between the noncollinear magnetic structure and ferroelectricity.

Single crystals of CuFe$_{1-x}$Al$_x$O$_2$($x$=0.00, 0.01, 0.02) were prepared by a floating zone method\cite{crystal}. For the measurements of pyroelectric current and dielectric constant, the crystals were cut into thin planes with the widest faces parallel to (1$\bar{1}$0) plane. As the electrodes, we painted silver paste onto these faces. Dielectric constant was measured at 100 kHz using an $LCR$ meter. For the electric polarization, we measured the pyroelectric current with a constant rate of temperature sweep and integrated it with time. Prior to the current measurements, a proper poling procedure was taken with electric field ($\sim$ 200 kV/m) perpendicular to the $c$-axis to obtain a single ferroelectric domain. In most cases, the current measurements were carried out at a rate of 5 $\sim$ 20 K/min in warming run once after cooling down to the lowest temperature with the poling electric field. In case the reentrant paraelectric behavior is observed, the poling was once stopped in the ferroelectric phase, and then the pyroelectric displacement current was measured with both increasing and decreasing temperature from the stopped temperature. AC and DC measurements of magnetization were done with a Physical Property Measurement System (PPMS, Quantum Design Inc.).
\begin{figure}
\begin{center}
\includegraphics*[width=8cm]{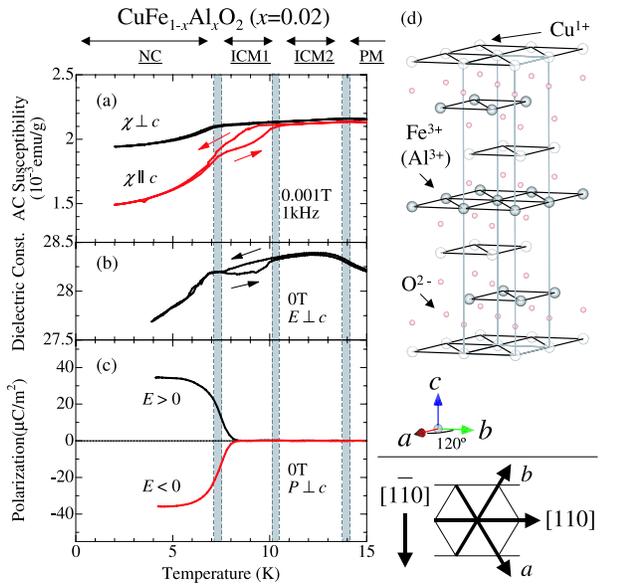}
\caption{(color online). Temperature profiles of (a) ac susceptibility measured with magnetic fields parallel and perpendicular to the $c$-axis, (b) in-plane dielectric constant, and (c) electric polarization perpendicular to the $c$-axis for CuFe$_{1-x}$Al$_x$O$_2$ ($x$=0.02). The arrows in (a) and (b) indicate the thermal scan direction of the measurement. Two opposite poling electric fields are used for the polarization measurement ($E>0$ and $E<0$). (d) Schematic crystal structure of delafossite CuFe(Al)O$_2$.}
\end{center}
\end{figure}

Figure 1 shows the temperature dependence of the AC susceptibility, dielectric constant, and polarization for $x$=0.02. There are known to be three magnetically ordered phases at zero magnetic field in this crystal\cite{x002}. While the wave vector of magnetic modulation is along (110) in all the phases, the direction of the magnetic moment is different among the phases. The sinusoidal and collinear magnetic structure shows up at 14 K (ICM2). While the magnetic moments are canted by 50 degree from the $c$-axis in the ICM2 phase\cite{OPD_structure}, they are aligned to the $c$-axis in the lower temperature phase below 11 K (ICM1)\cite{PD_structure}. The noncollinear magnetic structure (NC) is realized below 7 K. A broad peak of susceptibility around 14 K for both $H\parallel c$ and $H\perp c$ corresponds to the transition from paramagnetic phase (PM) to ICM2. Kinks in the susceptibility for $H\parallel c$ at 11 K and 7 K are caused by the transitions from ICM2 to ICM1 and from ICM1 to NC, respectively. The kink corresponding to the lower transition temperature is also observed in the susceptibility for $H\perp c$. The difference in the anisotropy of susceptibility between ICM1 and ICM2 is due to the direction of the magnetic moments. A large thermal hysteresis is observed in ICM1 in the susceptibility for $H\parallel c$. The dielectric constant shows kinks at 11 K and 7 K, similarly to the magnetic susceptibility. The hysteresis is also observed for the dielectric constant in the ICM1 phase. These suggest the strong coupling between the electric and magnetic properties in this material. Most importantly, the spontaneous polarization begins to increase at 7 K, which is the transition temperature of the noncollinear phase. We confirm the ferroelectric nature, that is the sign reversal of the polarization when the opposite poling field is used. This polarization is observed at zero magnetic field, in contrast with the case for $x$=0.00\cite{Kimura}, where polarization is induced only when magnetic fields of 6 - 13 T are applied along the $c$-axis.

\begin{figure}
\begin{center}
\includegraphics*[width=5cm]{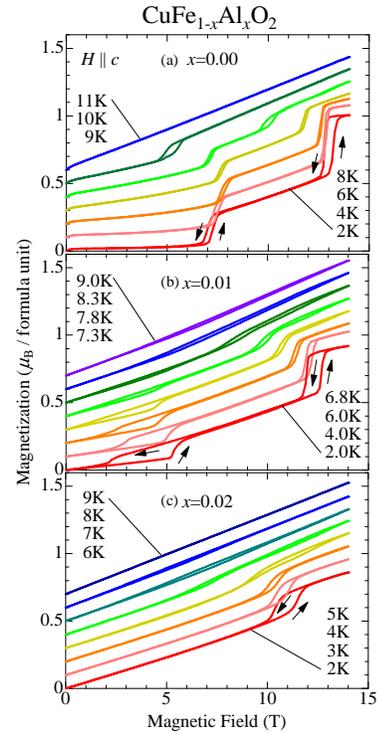}
\caption{(color online). Magnetic field dependence of magnetization at various temperatures for CuFe$_{1-x}$Al$_x$O$_2$ ((a)$x$=0.00, (b)$x$=0.01, (c)$x$=0.02). Arrows indicate the field-scan direction of the measurement. In the measurement for $x$=0.01, the sample was warmed up to 20 K and cooled down without magnetic field, prior to each field-increasing run.}
\end{center}
\end{figure}

Hereafter, we investigate the systematic evolution of the magnetoelectric phase with variation of Al-doping $x$. The magnetization curves for $x$=0.00 at various temperatures are shown in Fig. 2(a). Below 9 K, two notable steps are discerned in each curve. The step at the lower field corresponds to the transition from the commensurate 4 sublattice  (CM-4 : $\uparrow \uparrow \downarrow \downarrow $ with collinear spin directions along the $c$-axis) to the noncollinear magnetic structure. The higher-lying one is due to the emergence of the commensurate 5 sublattice  (CM-5 : $\uparrow \uparrow \uparrow \downarrow \downarrow $ with collinear spin directions along the $c$-axis) magnetic structure. The wave vectors of both CM-4 and CM-5 are also parallel to (110). Figure 3(a) depicts the temperature dependence of electric polarization at various magnetic fields for $x$=0.00. The spontaneous polarization is observed between 7 T and 12 T. This magnetic field region corresponds to the noncollinear magnetic phase. The results for $x$=0.00 are consistent with the previous report\cite{Kimura}.

\begin{figure}
\begin{center}
\includegraphics*[width=5cm]{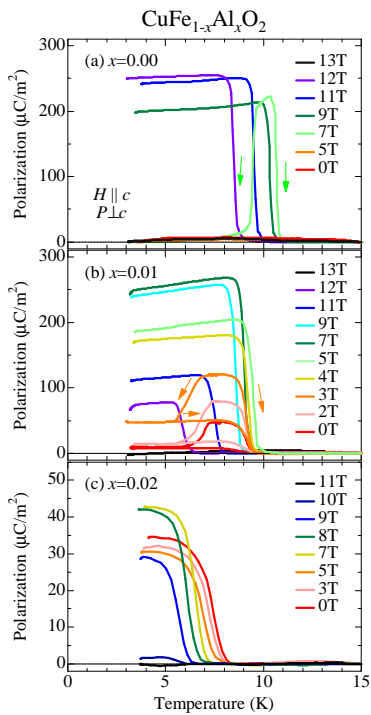}
\caption{(color). Temperature dependence of electric polarization perpendicular to the $c$-axis at various magnetic fields for CuFe$_{1-x}$Al$_x$O$_2$, (a)$x$=0.00, (b)$x$=0.01, (c)$x$=0.02. Arrows in (a) and (b) indicate the thermal scan direction of the measurement. After a proper poling procedure, each measurement was made in a warming run unless indicated by the arrows (see text).}
\end{center}
\end{figure}

In the case for $x$=0.01 (Fig. 2(b)), the two magnetic phase transitions are also observed below 7 K, while the transition field from CM-4 to NC considerably decreases. The larger magnetic-field hysteresis is observed for the transition than for $x$=0.00. Above 6.8 K, the magnetization is linear with the magnetic field below 7 T in a field-decreasing run, while the step-like structure is observed around 4 T in a field-increasing run (for the detail of the cooling procedure, see the figure caption). This suggests that NC phase exists even at zero field in this temperature region once after high enough field is applied. In Fig. 3(b), we show the temperature dependence of polarization at various fields for $x$=0.01. At zero field, the finite spontaneous polarization is observed between 6 K and 9 K in the first temperature scan after poling while it is almost negligible in a warming process after cooling down to 2 K without electric field. A similar behavior is observed below 3 T\cite{footnote3}. Above 4 T, there is a spontaneous polarization even at the lowest temperature. The polarization disappears at the NC - to - CM-5 transition field.

Figure 2(c) depicts magnetization curves for $x$=0.02. Only a single step is observed around 10 T below 4 K, as caused by the transition from the NC to CM-5 phase\cite{footnote1}. This confirms that the CM-4 phase is completely suppressed for the $x$=0.02 doping. Figure 3(c) shows the temperature dependence of polarization for $x$=0.02. The polarization begins to increase around 7 K with decreasing temperature at zero magnetic field. The reentrant paraelectric behavior as observed for $x$=0.01 at $H<3$ T is not observed for the case of $x$=0.02. The onset temperature of the spontaneous polarization decreases with magnetic field. The ferroelectric behavior disappears above 12 T. Thus, the ferroelectric region seems to coincide with the region of NC phase also for the Al-doped crystals.

The variations of these phase competitions with magnetic field and Al-doping are more clearly figured out by the magnetoelectric phase diagram (Fig. 4); the temperature versus magnetic field (perpendicular to the $c$-axis) phase diagrams for (a)$x$=0.00, (b)$x$=0.01, and (c)$x$=0.02. They were determined by the measurements of magnetization and dielectric constant. The phase diagrams for $x=0.00$ and $x=0.02$, apart from the identification of the ferroelectric state, agree with the previously reported ones determined by neutron scattering\cite{x000_FI,x002}. Although the neutron scattering was done only under zero magnetic field for $x$=0.01\cite{zero_field_diagram}, each magnetic phase can be specified by the comparison with the phase diagrams for $x$=0.00 and $x$=0.02. The ICM2 phase appears when Al is doped into CuFeO$_2$. The critical field for the transition from CM-4 to ferroelectric NC decreases dramatically but systematically with Al doping. The CM-4 phase is completely suppressed for $x$=0.02. The onset of the spontaneous polarization is also plotted with circles. Importantly, in each diagram, spontaneous polarization is observed only in the NC phase (shadowed region), not in the collinear nor paramagnetic phase. This indicates the firm relation between the noncollinear spin structure and ferroelectricity.

\begin{figure}
\begin{center}
\includegraphics*[width=5cm]{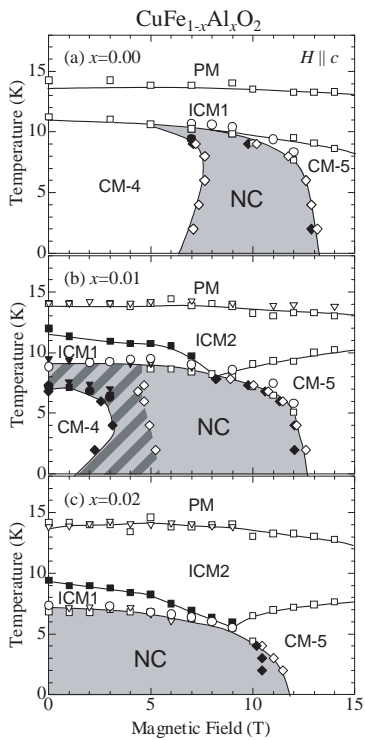}
\caption{Temperature ($T$) versus magnetic field ($H$) phase diagram of CuFe$_{1-x}$Al$_x$O$_2$ for (a)$x$=0.00, (b)$x$=0.01, and (c)$x$=0.02 ($H\parallel c$). Circle, triangle, square, and diamond data points were obtained by measurements of electric polarization, dielectric constant, magnetization ($T$-dependence), and magnetization ($H$-dependence), respectively. Open and filled symbols represent the anomalies in the increasing and decreasing $T$ or $H$, respectively. Spontaneous electric polarization was observed in the shadowed region. In the hatched area, 
the magnetic structure depends on the hysteresis. The CM-4 phase remains in a field-increasing or temperature-increasing run, otherwise the NC phase shows up in this area.}
\end{center}
\end{figure}

The presently observed ferroelectric NC phase may be relevant to the spin current model\cite{Katsura}, which has successfully explained the origin of ferroelectricity in some of recently found magnetic ferroelectrics. According to this model, the electric polarization $\vec{P}_{ij}$ produced between the two magnetic moments $\vec{S}_{i}$ and $\vec{S}_{j}$ is given by $\vec{P_{ij}} \propto \vec{e}_{ij} \times (\vec{S}_{i} \times \vec{S}_{j})$ with $\vec{e}_{ij}$ being the unit vector connecting the sites $i$ and $j$. In a noncollinear magnetic phase, in general, the term $\vec{S}_{i} \times \vec{S}_{j}$ shows a finite value. Therefore, the spin current model is consistent with the fact that the ferroelectricity is observed only in the noncollinear phase. Nevertheless, previous neutron diffraction studies\cite{x000_FI,ODSx002} suggest that the spin direction is always perpendicular to the wave vector in the noncollinear magnetic structures of both $x$=0.00 and $x$=0.02 crystals. This suggests that $\vec{S}_{i} \times \vec{S}_{j} \parallel \vec{e}_{ij}$, and hence contradicts with the spin current model. If this spin structure model for the incommensurate NC phase were correct, it would be very difficult to assign the origin of ferroelectricity for CuFe$_{1-x}$Al$_x$O$_2$ at the present stage. To revisit and possibly revise the magnetic structure may be needed\cite{meeting}.

In summary, we have demonstrated the emergence of ferroelectricity via the impurity (Al) doping ($x\sim 0.02$) in the triangular lattice antiferromagnet CuFeO$_2$. The magnetic phase diagram changes dramatically and in particular the critical magnetic field for the collinear - to - noncollinear (NC) transition decreases down to zero field with Al doping up to $x$=0.02. The finite electric polarization was observed only in the noncollinear magnetic structure, not in collinear one, suggesting the magnetic origin of the ferroelectricity. Thus, the modification of the spin frustration with site-dilution is another promising route to realize novel magnetic ferroelectrics.

The authors thank S. Miyasaka, N. Oda, S. Tanaka and H. Katsura for enlightening discussions. This work was partly supported by Grants-In-Aid for Scientific Research (Grant No. 15104006, 17340104, 16076205) from the MEXT of Japan.


\begin{thebibliography}{100}
\bibitem{meeting} Regarding the detail of the noncollinear magnetic structure, the previous neutron measurement suggested the proper screw structure for $x$=0.00\cite{x000_FI} and the double sinusoidal structure for $x$=0.02\cite{ODSx002}. However, the recent preliminary result suggests that the conclusion may need to be revised\cite{meeting2}.
\bibitem{footnote3}To be accurate, a small polarization remains even below the transition temperature of CM-4 phase for $x$=0.01 and increases with magnetic field. This is likely due to the phase coexistence of the CM-4 and NC associated with the first-order transition nature.
\end{thebibliography}
\end{document}